\documentstyle[psfig,prl,aps]{revtex}

\begin{document}
\draft

%%%Version of 8 September 1998

\def\pbar{\overline{p}}
\def\tpbar{\tau_{\overline{p}}}

%\pagestyle{empty}
%\vskip 0.5cm
\title{{\bf~\\\ \\\ \\\ \ \ \ \ \ \ \ \ \ 
THE COSMIC RAY ANTIPROTON SPECTRUM AND A 
LIMIT}\raisebox{0.75in}{\hspace{-1.5in}\parbox[t]
{2.1in}{\rm FERMILAB-PUB-98/265-A\ \ \\ UF-IFT-HEP-98-13}}
{\bf\ \ ON THE ANTIPROTON LIFETIME}}
%\vskip 1.0cm
\author{{\bf Stephen H. Geer}$^{*}$}
%\vskip 0.5cm
\address{\it Fermi National Accelerator Laboratory, P.O. Box 500, 
Batavia, Illinois 60510 USA}
%\vskip 1.0cm
\author{{\bf Dallas C. Kennedy}$^{**}$}
%\vskip 0.5cm
\address{\it Department of Physics, University of Florida, 
Gainesville, Florida 32611 USA}
%\vskip 0.75cm
%\medskip
\date{\today}
\maketitle
\begin{abstract}
%\vskip 0.5cm
Measurements of the cosmic ray $\pbar /p$ ratio are 
compared to predictions from an inhomogeneous leaky box model of Galactic 
secondary $\pbar$ production combined with an updated heliospheric model 
that modulates the predicted fluxes measured at the Earth. The heliospheric 
corrections are crucial for understanding the low--energy part of the 
spectrum. The $\pbar$ production-propagation-modulation model agrees with 
observations. Adding a finite $\pbar$ lifetime to the model, we obtain 
the limit $\tpbar >$ 1.7 Myr (90\% C.L.).
Restrictions on heliospheric properties and the cutoff of 
low-energy interstellar cosmic rays are discussed.
%\vskip 1.0cm
\end{abstract}

\pacs{11.30.Er, 14.20.-c, 95.30.Cq, 96.40.-z, 96.40.Cd, 98.70.-f}
\centerline{Submitted to {\bf Physical Review Letters}}
\bigskip

%\vfill\eject
%\pagestyle{plain}

In recent years the presence of antiprotons $(\pbar$'s) in the 
cosmic ray (CR) flux incident upon the Earth has been firmly 
established by a series of balloon--borne 
experiments~\cite{golden,bog1,bog2,bog3,hof,labrador,moiseev,yoshimura,boezio}. 
The CR $\pbar /p$ ratio has been measured from kinetic energies
of $\sim$ 100 MeV to $\sim$ 19 GeV. The measurements are summarized 
in Table~1. The CR $\pbar /p$ spectrum can be predicted using the Leaky 
Box Model (LBM)~\cite{steph,webbpot,gaisssch}, which assumes 
that the $\pbar$'s originate from proton interactions in the 
interstellar (IS) medium. The $\pbar$'s then propagate within the 
Galaxy until they ``leak out" with the characteristic 
CR Galactic storage time of about 10 million years (Myr)~\cite{galmag}.  
To obtain a prediction for the energy-dependent $\pbar /p$ ratio at 
the Earth, the IS spectrum must be corrected for modulation of the 
$p$ and $\pbar$ fluxes as the particles propagate through the 
heliosphere. Thus, a comparison of the observed $\pbar /p$ spectrum 
with predictions tests the Galactic CR production-propagation model
and the relevant heliospheric parameters. In addition, if the 
$\pbar$'s have a lifetime $\tpbar$ short compared to the 
Galactic storage time, the observed $\pbar$ spectrum will be 
depleted~\cite{golden,bog1,steph}, 
particularly at low energies. This facilitates an interesting test of 
CPT invariance, which requires $\tpbar$ = $\tau_p$, where the
proton lifetime $\tau_p$ is known to exceed
${\cal O}(10^{32})$ yr~\cite{ppbarprops}.

In this paper the observed CR $\pbar /p$ spectrum is shown to be 
well--described by the predictions of an improved variant of the LBM, once
modulation corrections are included.  The data constrain the acceptable
heliospheric properties.  Extending the LBM to permit an unstable $\pbar$,
we obtain a limit on $\tpbar$ significantly more stringent than current 
laboratory bounds obtained from searches for $\pbar$ decays in ion
traps~\cite{antiatom} and storage rings~\cite{apex}.

In the LBM the only source of IS $\pbar$'s is the 
spallation of CR $p$'s~\cite{steph,webbpot,gaisssch}: 
$pN_Z\rightarrow\pbar X,$ where $N_Z$ is a nucleus of charge $Z$, and $X$ is 
anything.  
Realistic elemental abundances of the nuclei in the IS medium are given in 
refs.~\cite{webbpot,gaisssch}. 
The cross section for $Z$ = 1 (the dominant contribution) has been measured
in accelerator experiments~\cite{steph}.
For $Z >$ 1, some assumptions about nuclear
physics are required to make use of the $Z$ = 1 data.  We have used the 
``wounded nucleon'' picture of~\cite{gaisssch}.
The $\pbar$'s are assumed to propagate within the Galaxy 
until they are lost by one of two processes.  
The dominant loss process is leakage into intergalactic space. This process 
also applies to other CRs, including protons. 
The storage lifetime associated with this leakage mechanism 
varies with rigidity $P$ and can be inferred from measurements 
of unstable secondary CR nuclei with known decay halflives.
The sub--dominant $\pbar$ loss mechanism is $p\pbar$ annihilation. 
The annihilation rate is considerably smaller than the leakage rate.  
Leakage and annihilation, together with elastic and inelastic scattering 
of $\pbar$'s, are the ingredients of the {\it standard
leaky box model} (SLBM), in which production and loss are assumed to be 
in statistical equilibrium. 

Our analysis is based on the SLBM of Gaisser and Schaeffer 
using the parameters of \cite{gaisssch}, but with a 
Galactic storage time improved to
account for the non-uniform Galactic CR distribution~\cite{galmag}. 
This {\em inhomogeneous LBM} (ILBM) results in a somewhat longer storage time,
$4.1\times 10^{14}~[1+P/(3{\rm\ GeV})]^{-0.6}$ sec.
The uncertainties on the parameters of the ILBM result in uncertainties on 
the normalization of the predicted $\pbar /p$ ratio but, to a good 
approximation, 
do not introduce significant uncertainties in the shape of the predicted 
spectrum. Specifically, uncertainties on four ILBM parameters must be 
considered: (i) the storage time ($\pm$67\%~\cite{galmag}), 
(ii) the IS primary $p$ flux ($\pm$35\%~\cite{gaisssch}), 
(iii) the $\pbar$ production cross section ($\pm$10\%~\cite{steph,gaisssch}), 
and (iv) the composition of the
IS medium, which introduces an uncertainty of $<6$\% on the predicted $\pbar$ 
flux~\cite{gaisssch}. 
In the following we neglect the last of these uncertainties since 
it is relatively small. 
Within the quoted fractional uncertainties on the other three parameters we 
treat all values as being a priori equally likely. Note that the predicted 
$\pbar /p$ ratio is approximately proportional to each of the 
three parameters under consideration. 

The curves in Fig.~1 show the ILBM $\pbar /p$ 
spectra for the parameter choices that result in the largest and smallest 
$\pbar /p$ predictions. The predicted IS spectrum does not give 
a good description of the observed spectrum at the top of the atmosphere. 
Good agreement is not expected because the CR spectra observed at the Earth 
are modulated as the particles propagate into the Sun's magnetic region of 
influence, the heliosphere~\cite{solsys,cosmics}, which consists of
the solar magnetic field ${\bf B}$ and the solar wind.  
Measurements by the {\it Pioneer 10} and {\it Voyager 1}
spacecraft~\cite{pionvoy} indicate 
that the outer boundary of the heliosphere (the heliopause) is at 
heliocentric distance $r_0\geq$ 71 astronomical
units (AU). Theoretical considerations favor $r_0$ =
100--130 AU~\cite{solsys}.  Fortunately, the heliospheric corrections 
to the IS fluxes are not sensitive to variations of $r_0$ within these 
ranges.
The wind, which is assumed to blow radially outwards, has an equatorial speed 
$V_W\sim$~400~km~sec$^{-1}$.  Away from the equatorial plane 
the {\it Ulysses} spacecraft has found $V_W\sim$~750~km~sec$^{-1}$~\cite{ulys}. 
The wind pressure declines until it reaches the IS value at the heliopause. 
The solar wind plasma (nuclei and $e^-)$ moves in 
bulk outwards from the Sun with turbulence that peaks around
solar magnetic maximum. 
The solar wind imposes an energy threshold for IS CRs reaching the Earth. 
This energy cutoff is much higher for particles coming in along the polar 
directions where the wind speeds are higher, than 
along the equatorial route where the wind speeds are lower.
The wind also carries the solar magnetic 
flux outward. Solar rotation $\Omega_\odot$ twists the field lines 
to form a Parker spiral. 
The smoothed heliomagnetic field 
($B_\oplus\sim$ 5 nT at the Earth's orbit) declines as it
changes from radial at the Sun to azimuthal in the outer Solar System.  
The heliomagnetic polarity (${\rm sign}(A)$)
is opposite in northern and 
southern solar hemispheres and switches sign somewhat after sunspot maximum
(roughly every 11 years), when the field becomes more disordered.
The regions of opposite magnetic polarity are separated by an approximately
equatorial, unstable neutral current sheet.  The sheet is wavy and spiraled;
its waviness is measured by its ``tilt'' angle $\alpha$, 
which relaxes from $\simeq$ $50^\circ$ at polarity reversal to $\lesssim$
$10^\circ$ just before reversal~\cite{solsys,ulys,sunspot,transport}.

Cosmic rays enter the heliosphere on ballistic trajectories 
and are then subject to the forces associated with the wind and solar magnetic 
field~\cite{cosmics,transport}. The propagation of the CRs within the 
heliosphere is described by a drift-diffusion (Fokker-Planck) 
equation~\cite{fpeqn}.  The CRs are pushed outwards
by the bulk motion of the wind (elastic scattering), lose energy
as they perform work on the wind
(adiabatic deceleration or inelastic scattering), are
diffusively scattered by field turbulence, and
execute a drift {\it orthogonal} to the curving magnetic
field lines as they spiral inwards {\it along} the
field lines.  Particles with $qA > 0$ $(< 0)$ drift in along a polar 
(sheet) route.
The IS particles with sufficient energy to overcome the various energy losses
reach the inner Solar System after being randomly scattered by 
turbulence. 

We compute the modulation of CR fluxes by the method of characteristics and
combined Runge-Kutta/Richardson-Burlich-Stoer techniques~\cite{diffeqns}.  
The calculation uses the heliospheric transport 
models of Jokipii {\it et al.}~\cite{transport}, combined with the
ILBM $\pbar$ spectrum, and updated by {\it Pioneer, Voyager,} and 
{\it Ulysses} heliospheric measurements~\cite{pionvoy,ulys}.  
Our calculation includes magnetic curvature drift since older heliospheric 
models~\cite{transportOLD} that neglected this drift component
have been shown~\cite{transport} to be inadequate. 
Note that magnetic drift changes sign for oppositely-charged
particles and field polarities, in contrast to diffusion $\kappa$ which is 
charge-invariant and whose effect partly cancels in the $\pbar /p$ ratio. 
%Drift and deceleration shift particles in phase space; turbulence reduces
%their number at the heliopause and superposes a random walk on survivors
%with a velocity $v$ described by a mean free path $\kappa /v$.
The calculation is simplified by 
ignoring turbulence where its effects are small, 
which in practice means everywhere except across the sheet, where for particles
with speed $v$, we use:
$\kappa_\perp$ = $(2-3)\times 10^{17}[B_\oplus /B(r)](P/{\rm GeV})^{0.3}(v/c)$
m$^2$/sec~\cite{elsewhere}.
Finally, IS ``pickup'' ions and solar CRs have been 
neglected~\cite{solsys,ulys}.

In the following, we restrict our analysis to the data sets
recorded by the MASS91, IMAX, BESS, and CAPRICE experiments 
(Table~1). 
These data were recorded in the period 1991--1994, corresponding to a 
well--behaved part of the solar cycle
for which the heliospheric modulation corrections 
can be calculated with some confidence.
In addition, compared to the earlier
data, these more recent data sets benefited from
detectors with improved particle identification
and hence lower backgrounds. 
In Fig.~2 the measurements are
compared with the predictions for a stable antiproton,
taking the central parameter values for our ILBM
calculation and three different sets of heliospheric corrections (lower, 
central, and upper). 
The corresponding heliospheric parameters are 
(a) Lower: equatorial $V_W =$~375~km~sec$^{-1}$, 
polar $V_W =$~700~km~sec$^{-1}$, 
$B_\oplus =$~4.0~nT;
(b) Central: equatorial $V_W =$~400~km~sec$^{-1}$, 
polar $V_W =$~750~km~sec$^{-1}$, 
$B_\oplus =$~4.5~nT; and 
(c) Upper: equatorial $V_W =$~425~km~sec$^{-1}$, 
polar $V_W =$~800~km~sec$^{-1}$, 
$B_\oplus =$~5.0~nT.
Note that for each parameter
set there are four curves, corresponding to the calculated
heliospheric corrections at the dates of each of the four
balloon flights. 
Given the variation of the predicted ratios with heliospheric 
parameter choice, we conclude that the predictions are able to 
give a good
description of the data, and that there is no significant
evidence for a short antiproton lifetime.
Note that
(i) for a given set of parameters, the heliospheric
corrections are similar for the BESS, IMAX, CAPRICE, and
MASS91 data, with appreciable differences only at the lowest
kinetic energies; and
(ii) in the region above $\sim 3$~GeV, the differences
between the predictions corresponding to the different choices
of heliospheric parameters are small compared to the
statistical errors on the data. However, in the lowest energy
region, below $\sim700$~MeV, there are significant differences
in the predictions for the three different choices of heliospheric
parameters. Hence, in the fitting procedure employed to extract
an upper limit on $\tpbar$, we must allow 
the overall normalization of the predictions to vary to account
for the uncertainties arising from the choice of the ILBM parameters,
and the heliospheric parameters to vary to
account for the sensitivity of the low-energy $\pbar /p$ ratio 
to these parameters.

To obtain a limit on $\tpbar$ we add to the ILBM one additional loss
mechanism, $\pbar$ decay.
Note that the $\pbar$ lifetime must be $< {\cal O}(10)$~Myr to significantly 
distort the $\pbar$ spectrum.
However, this timescale is much longer than the lifetime sensitivity
of existing laboratory searches for $\pbar$ decay~\cite{apex}.
The results from maximum likelihood fits to the BESS, IMAX, CAPRICE,
and MASS91 measurements are shown in the inset of Fig.~2
as a function of the assumed $\tpbar$ for eight heliospheric parameter sets 
F1 -- F8. The F1 parameters are equatorial (polar) 
$V_W$ = 375 (700)~km~sec$^{-1}$, and $B_\oplus$ = 4.0~nT. The 
values for the equatorial (polar) $V_W$ and $B_\oplus$ increase for the 
other parameter sets in steps of respectively 5 (10) km sec$^{-1}$ and 0.1 nT. 
Hence, the parameters for F8 are equatorial (polar)
$V_W$ = 410 (770)~km~sec$^{-1}$, and $B_\oplus$ = 4.7~nT. 
The fits, which take account of the Poisson
statistical fluctuations on the number of observed events and the
background subtraction for each data set (Table~1), also 
allow the normalization of the ILBM predictions to vary within 
the acceptable range (Fig.~1). 
The best maximum likelihood fits correspond to the heliospheric 
parameter sets F4 and F5, and are consistent 
with a completely stable antiproton. The extreme
parameter sets (F1 and F8) are excluded at $>90$\%~C.L. for all values 
of $\tpbar$, and hence the data constrain the heliospheric parameters. 
In particular, the following parameter ranges are preferred: 
equatorial $V_W$ = 375--410~km sec$^{-1}$, 
polar $V_W$ = 700--770~km sec$^{-1}$, 
and $B_\oplus$ = 4.0--4.7~nT.
Finally, the fits do not favor values of $\tpbar < {\cal O}(1)$ Myr. 
We obtain the bounds:
\begin{equation}
\tpbar\quad >\quad 1.7\ {\rm Myr\ \ (90\%\ C.L.)}\quad ,\quad
                   1.1\ {\rm Myr\ \ (95\%\ C.L.)}\quad ,\quad
                   630\ {\rm kyr\ \ (99\%\ C.L.)}\quad .\nonumber \\
\end{equation}

A new generation of balloon experiments, or a dedicated orbital CR detector 
such as the recently tested AMS~\cite{ams}, could greatly improve
the statistical 
precision of the measured $\pbar$ spectrum.  This 
would facilitate more significant bounds on the heliospheric parameters, 
and enable a search for exotic sources of $\pbar$'s, particularly at low 
kinetic energies $K$. However we note that, neglecting diffusion, the solar 
wind cuts off the polar--routed flux at $K_\oplus$ = 0.26--0.40 GeV  
(mass-independent
cutoff rigidity scaling approximately as $qB_\oplus V^2_W$). 
Particles with these threshold energies at the Earth have IS energies 
$K_{\rm IS}$ = 0.54--0.80~GeV. In 
contrast, the sheet-routed flux exhibits no significant cutoff. 
Noting that in the present solar cycle the protons reach the Earth via 
the polar route, and $\pbar$'s reach the Earth via the sheet route, 
we conclude that when the next solar cycle begins (around 2003) 
new experiments will only be sensitive to IS 
$\pbar$'s with $K_{\rm IS} \gtrsim 0.5$~GeV. 
%Diffusion would relax the cutoff somewhat.
%The heliosphere also modulates the flow of
%other charged particles, such as $e^\pm$, nuclei,
%dark matter CHAMPs, and magnetic monopoles.

In summary, with an inhomogeneous LBM of IS $\pbar$ production and 
propagation, together with a calculation of heliospheric corrections 
that includes up--to--date data on the solar wind, we obtain good 
agreement with the observed CR $\pbar /p$ ratio. Extending the ILBM 
model to permit a finite $\tpbar$ we obtain lower limits on $\pbar$ 
decay that are more stringent than laboratory bounds. Our fits also 
constrain the ranges of the heliospheric parameters. We note that 
in the next solar cycle the effective cutoff in detectable low 
energy IS $\pbar$'s will be significantly higher than at present, 
which will decrease the sensitivity of experiments searching for exotic 
sources of antiprotons.

% THE FOLLOWING PARAGRAPH WILL EITHER BE ELIMINATED, OR IF WE HAVE SPACE,
% ITS CONTENT WILL BE PARTIALLY OR COMPLETELY ABSORBED INTO THE INITIAL 
% PARAGRAPHS
%Antiprotons are the only antimatter
%capable of decaying into purely known particles.  A finite $\pbar$
%lifetime is therefore of intrinsic interest, as well as of interest in
%the context of baryogenesis.  Furthermore, a $\pbar$ lifetime {\it 
%different} from the $p$ lifetime implies a violation of CPT invariance,
%which requires particle and antiparticle properties (mass, $|$charge$|$, 
%lifetime)
%to be the same, a central result of special relativity and
%quantum mechanics, along with the related spin-statistics 
%theorem~\cite{cptthm}.  It holds under several broad assumptions of field 
%theory (Lorentz invariance, microcausality, locality, and uniqueness
%of the vacuum) and, with conventional compactifications, in
%superstring theories~\cite{cptstrings}.  These assumptions can be violated, 
%however, in exotic but not impossible scenarios in field and string
%theory~\cite{cptrecent} or outside of ordinary quantum 
%mechanics~\cite{qmviol}.  The CPT symmetry is thus tested by antiproton decay
%searches.  Other sensitive tests of the CPT symmetry are
%based on laboratory-produced positrons and antiprotons~\cite{cpttests} and 
%$K^0-\overline{K^0}$ neutral meson mixing and decay~\cite{kkbar}, as well as
%neutron interferometry and solar neutrinos~\cite{qmviol}.***

\bigskip

\noindent{\bf Acknowledgements}

\bigskip
It is a pleasure to thank Thomas Gaisser, 
%(Bartol Research Institute), 
J.~R.~Jokipii, 
%(Lunar \& Planetary Laboratory), 
and E.~J.~Smith 
%(Jet Propulsion Laboratory) 
for their insights.
This work was supported at Fermilab 
%and Theoretical Astrophysics Group 
under grants U.S. DOE DE-AC02-76CH03000 and NASA NAG5-2788.
%respectively.  D.C.K. also thanks 
and at the U. Florida, Institute for Fundamental
Theory, under grant U.S. DOE DE-FG05-86-ER40272.

%\vskip 0.5cm

\clearpage
\thispagestyle{plain}
\textheight 8.5in
\begin{table}
\begin{center}
\begin{tabular}{|lc|c|c|c|c|c|c|c|} 
Experiment& &Field&Flight&KE Range&Cand-&Back-&Observed &predict-\\
          & &Pol.&Date&(GeV)&idates&ground&$\overline{p}/p$ Ratio&tion${}$\\
\hline
Golden$^\dag$&\cite{golden}&$+$&June 1979&5.6 -- 12.5&46&18.3&$(5.2\pm 1.5)\times 10^{-4}$&
- \\ 
\hline
Buffington$^\dag$&\cite{buff}&${\rm n/a}$&June 1980&0.13 -- 0.32&17&3.0&$(2.2\pm 0.6)
\times 10^{-4}$& - \\ 
\hline
Bogomolov$^\dag$&\cite{bog1}&$+$&1972-1977&2.0 -- 5.0&2&-&$(6\pm 4)\times 10^{-4}$& 
- \\
Bogomolov$^\ddag$&\cite{bog2}&$-$&1984-1985&0.2 -- 2.0&1&-&$(6^{+14}_{-5})\times 
10^{-5}$& - \\
Bogomolov$^\ddag$&\cite{bog3}&$-$&1986-1988&2.0 -- 5.0&3&-&$(2.4^{+2.4}_{-1.3})\times 
10^{-4}$& - \\ \hline
PBAR$^\dag$&\cite{ahlen}&$-$&Aug. 1987&0.205 -- 0.64&n/a&n/a&$\leq 4.6\times 10^{-5}$&
- \\
PBAR$^\dag$&\cite{salamon}&$-$&Aug. 1987&0.10 -- 0.64 &n/a&n/a&$\leq 2.8\times 10^{-5}$&
- \\
PBAR$^\dag$&\cite{salamon}&$-$&Aug. 1987&0.64 -- 1.58 &n/a&n/a&$\leq 6.1\times 10^{-5}$&
- \\
\hline
LEAP$^\dag$&\cite{streit}&$-$&Aug. 1987&0.12 -- 0.86&n/a&n/a&$\leq 1.8\times 10^{-5}$&
- \\
\hline
MASS91&\cite{hof}&$+$&Sep. 1991&3.70-19.08&11&3.3&$(1.24^{+0.68}_{-0.51})\times 
10^{-4}$&$1.3\times 10^{-4}$\\
\hline
IMAX&\cite{labrador}&$+$&July 1992&0.25 -- 1.0&3&0.3&$(3.14^{+3.4}_{-1.9})\times
10^{-5}$&$1.2\times 10^{-5}$\\
IMAX&\cite{labrador}&$+$&July 1992&1.0 -- 2.6 &8&1.9&$(5.36^{+3.5}_{-2.4})\times
10^{-5}$&$6.2\times 10^{-5}$\\
IMAX&\cite{labrador}&$+$&July 1992&2.6 -- 3.2 &5&1.2&$(1.94^{+1.8}_{-1.1})\times
10^{-4}$&$1.1\times 10^{-4}$\\
\hline
BESS&\cite{moiseev}&$+$&July 1993&0.20 -- 0.60&6&$\sim 1.4$&$(5.2^{+4.4}_{-2.8})\times 10^{-6}$&
$6.4\times 10^{-6}$\\
\hline
CAPRICE&\cite{boezio}&$+$&Aug. 1994&0.6 -- 2.0 &4&1.5&$(2.5^{+3.2}_{-1.9})\times 
10^{-5}$&$3.3\times 10^{-5}$\\
CAPRICE&\cite{boezio}&$+$&Aug. 1994&2.0 -- 3.2 &5&1.3&$(1.9^{+1.6}_{-1.0})\times 
10^{-4}$&$1.1\times 10^{-4}$\\
\end{tabular}
\vspace{0.5cm}
\caption{
Summary of cosmic ray antiproton results.  Listed from left to right 
are the experiments, solar cycle polarity (note: 
minimum-to-minimum solar cycles 20:1964.6--1976.5, 21:1976.5--1986.7, 
and 22:1986.7--1996.3), balloon flight date, $\pbar$ kinetic energy range, 
number of $\pbar$ candidates observed, estimated number of background events, 
measured $\pbar /p$ ratio at the top of the atmosphere, and the ILBM 
prediction using the 
heliospheric parameters: equatorial (polar) $V_W$ = 395 (740)~km~sec$^{-1}$, 
$B_\oplus$ = 4.4~nT.
\dag~Not shown in Fig.~1 or used in analysis.  
\ddag~Not used in analysis.
}
\label{balloon_tab}
\end{center}
\end{table}

\begin{figure*}[tl]
%\leavevmode\centering
\psfig{file=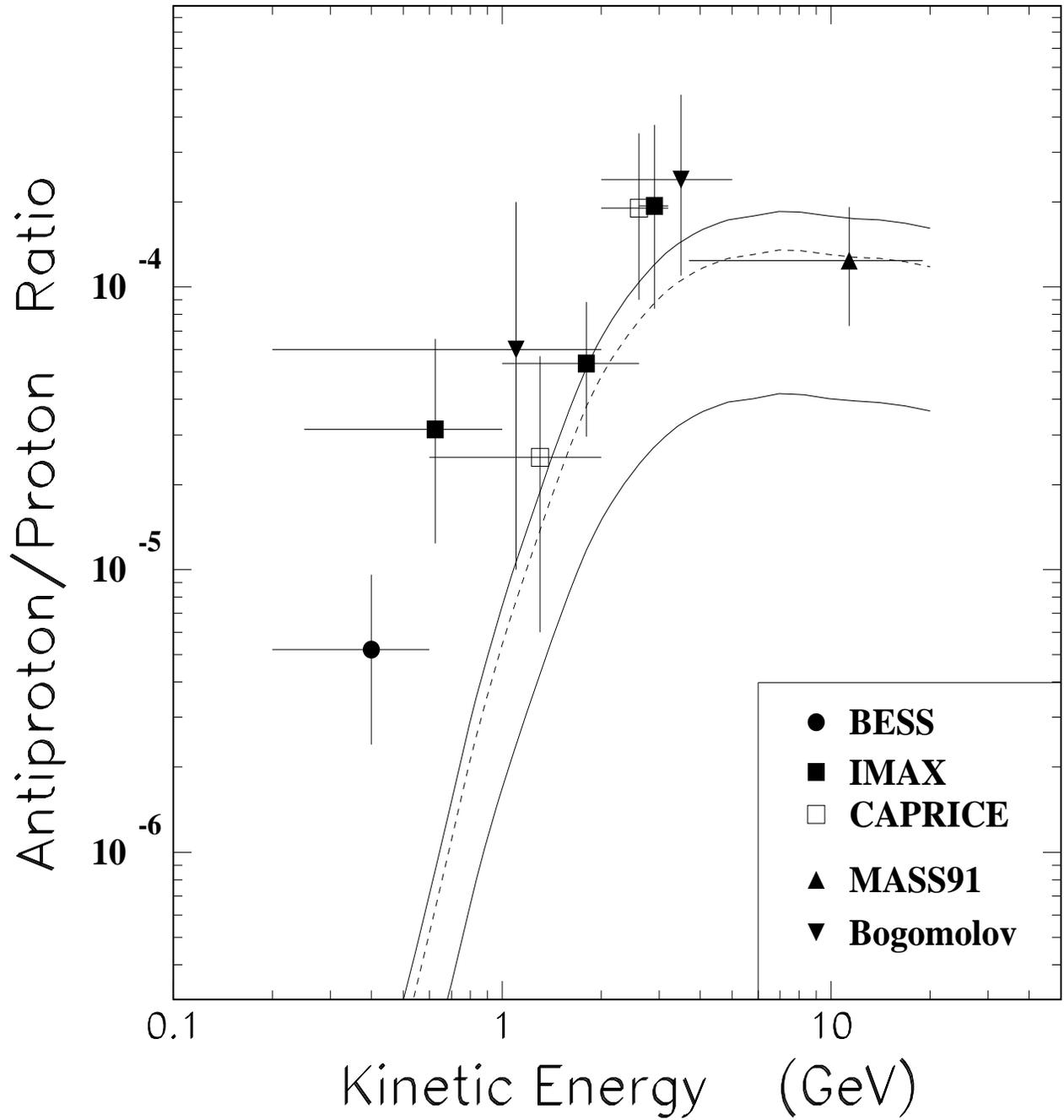,width=6.8in,height=7.2in}
\ \\\ \\
\caption{
Observed $\pbar /p$ ratio at the top of Earth atmosphere (see Table~1).  
The solid curves show the upper and lower interstellar ratios 
predicted by the ILBM described in the text, without solar modulation. 
The broken curve shows the ILBM prediction with the same parameters 
used for the modulated predictions of Fig.~2.}
\label{fig1}
\end{figure*}

\begin{figure*}[tl]
\leavevmode\centering\psfig{file=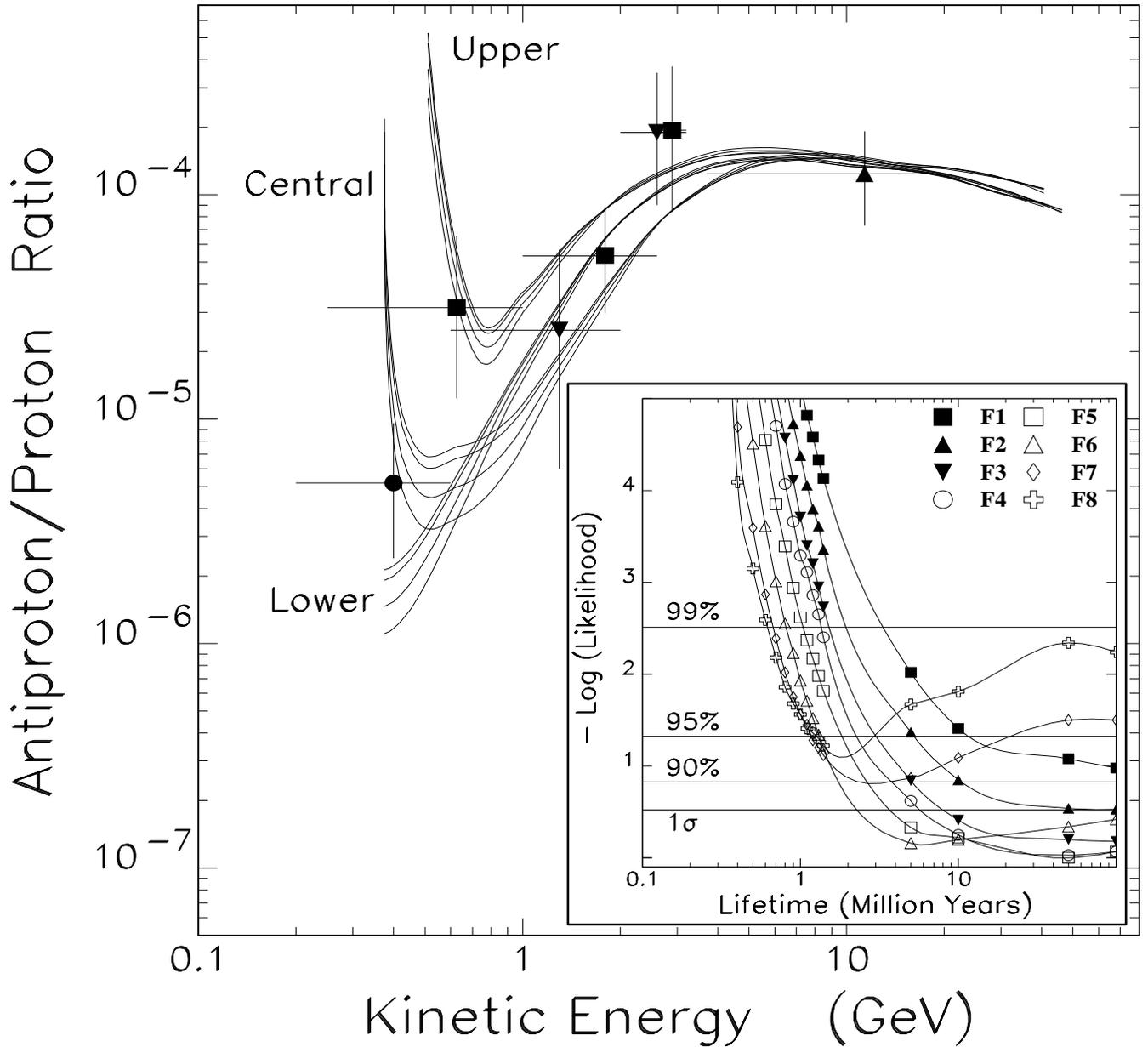,width=7.5in,height=7.0in}
\ \\\ \\\ \\\ \\\ \\
\caption{
Observed $\pbar /p$ ratio at the top of Earth atmosphere compared with 
the ILBM predictions (see Fig.~1) after heliospheric modulation. 
The upper, central, and 
lower families of curves correspond to the upper, central, and lower 
heliospheric parameter sets described in the text. For each heliospheric 
parameter set there are four curves corresponding to the epochs of the 
four balloon flights (at low energies, from top to bottom: MASS91, IMAX,
BESS, CAPRICE).
The inset shows the fit results as a function of the assumed $\tpbar$ 
for the eight heliospheric parameter sets (F1 -- F8) described in the 
text.}
\label{fig2}
\end{figure*}


\begin{thebibliography}{99}
\bibitem{elsewhere}
Note that $\kappa_\perp$ would be smaller around solar maximum.
\end{thebibliography}
\end{document}